\documentclass[12pt]{article}

\usepackage{amsmath}
\usepackage{amssymb}

\usepackage[utf8]{inputenc}
\usepackage[T1]{fontenc}

\def\dju{\mbox{Đurđevich}}

\newtheorem{theorem}{Theorem}[section]
\newtheorem{lemma}{Lemma}[section]
\newtheorem{corollary}{Corollary}[section]
\newtheorem{prop}{Proposition}[section]
\newtheorem{definition}{Definition}[section]

\numberwithin{equation}{section}

\begin{document}
\begin{center}
{\bf Toeplitz Quantization for \\ Non-commutating Symbol \\ Spaces such as $SU_q(2)$} 
   \vskip 0.4cm
   Stephen Bruce Sontz
   \\
   Centro de Investigaci\'on en Matem\'aticas, A. C.
   \\
   (CIMAT)
   \\
   Guanajuato, Mexico
   \\
   email: sontz@cimat.mx
   \vskip 0.4cm
   Dedicated to the memory of Jaime Cruz Sampedro
\end{center}

\begin{abstract} 
Toeplitz quantization is defined in a general setting in which the symbols are
the elements of a possibly {\em non-commutative} algebra
with a conjugation and a possibly {\em degenerate} inner product. 
We show that the
quantum group $SU_q(2)$ is such an algebra. 
Unlike many quantization schemes,  
this Toeplitz quantization does not require a {\em measure}. 
The theory is based on the mathematical structures defined and 
studied in several recent papers of the author; those  papers 
dealt with some specific examples of this new Toeplitz quantization. 
Annihilation and creation operators are defined
as densely defined Toeplitz operators acting in a quantum Hilbert space, 
and their commutation relations are discussed.
At this point Planck's constant is introduced into the theory.
Due to the possibility of {\em non-commuting} symbols, there are now two definitions
for anti-Wick quantization; 
these two definitions are equivalent in the commutative case.
The Toeplitz quantization introduced here 
satisfies one of these definitions, but not necessarily the other.
This theory should be considered as a second quantization, since 
it quantizes {\em non-commutative} (that is, already {\em quantum}) objects. 
The quantization theory presented here has two essential features of a 
physically useful quantization: Planck's constant and a Hilbert space 
where natural, densely defined operators act. 
\end{abstract}


\section{Introduction}

The history of Toeplitz operators covers a bit over one hundred years and includes many major
works, far too numerous to mention here. 
For a recent reference that will give the reader some first links 
to that extensive literature, see Section~3.5 in \cite{ruben}.
Speaking for myself, the papers \cite{berger-coburn1}, \cite{berger-coburn2} and \cite{brian} have been
rather influential. 
But the study of Toeplitz operators with symbols coming from a
non-commutative algebra  seems to be limited mostly 
to cases where the algebra
is a matrix algebra or is some other quite specific non-commutative algebra 
such as in two recent works,  \cite{part2} and \cite{quantum-plane}, of the author. 
The papers \cite{part2} and \cite{quantum-plane} can be considered as two rather 
elaborated examples of the theory presented here. 
That study is continued in this paper, but in a much more general setting intended to
clarify the mathematical structures at play in those two examples.
A new example, the quantum group $SU_q(2)$ as symbol space, will also be presented here. 

The paper is organized as follows.
After presenting the foundations of this theory in the next section, we define and analyze in 
Section~\ref{TQ-section} the Toeplitz quantization. 
In particular, Toeplitz operators are defined as densely defined operators acting in a
quantum Hilbert space.
The symbols of these Toeplitz operators come from a possibly {\em non-commutative} algebra 
$\mathcal{A}$, which in physics terminology serves as the phase space for the theory. 
The common domain of these Toeplitz operators is $\mathcal{P}$, a pre-Hilbert space and sub-algebra 
of $\mathcal{A}$. 
$\mathcal{P}$ represents the possibly {\em non-commuting holomorphic polynomials}
in $\mathcal{A}$. 
We also discuss another important additional structure, namely, a 
projection operator $P : \mathcal{A} \to \mathcal{P}$. 
We emphasize that in order to quantize a given algebra $\mathcal{A}$ with conjugation 
we need to choose appropriately $\mathcal{P}$ and $P$. 
There is also a choice to be made of a possibly {\em degenerate} inner product on $\mathcal{A}$. 
As examples make clear, these choices are not unique. 
And in general there is no reason as far as I know to suppose that a given algebra $\mathcal{A}$ 
must necessarily have these extra structures. 
So that remains a question for future inquiry. 

In Section~\ref{section-seeking-P} we discuss 
how to find a sub-algebra 
$\mathcal{P}$ of $\mathcal{A}$ to represent the possibly {\em non-commuting holomorphic polynomials}. 
This technical point 
is nonetheless central to the application of the theory. 
In Section~\ref{CA-section} we define two special cases of Toeplitz operators, 
the creation and annihilation operators.
Then next in Section~\ref{AW-quantization-section} 
we study the relation of this Toeplitz quantization with two plausible 
definitions of anti-Wick quantization, each of which arises naturally in a non-commutative setting.
(These two definitions turn out to be equivalent in the commutative setting.) 
This Toeplitz quantization always satisfies one of these definitions, which is of course then taken
to be the `correct' generalization of anti-Wick quantization for this setting.
The notion of canonical commutation relations in this abstract context is discussed in
Section~\ref{CCR-section}. 
At this point Planck's constant is introduced into the quantum side of the theory. 
In Section~\ref{SUq2-section} we show that this theory can be 
applied in the setting of non-commutative geometry. 
We define and study a Toeplitz quantization whose symbols lie in the 
quantum group $SU_q(2)$ and which acts (possibly via unbounded operators)
on the Manin quantum plane, viewed as a pre-Hilbert space 
with respect to a specific inner product. 
In the last section we conclude with a few remarks about our approach to
Toeplitz quantization with non-commuting symbols as contrasted with other approaches. 

One notable feature, one might even say failing if one were unkind, 
of this theory is that almost none of the standard 
structures of a phase space is imposed on $\mathcal{A}$. 
For example, measures, symplectic forms, Poisson brackets and coherent states 
are not needed, though some of these structures could 
be present in some examples. 
The most important structure on $\mathcal{A}$ is strangely enough 
the conjugation operation. 
Even the multiplication on $\mathcal{A}$ (as an algebra) is not a critically  
important structure and can be
dispensed with as is done in \cite{toeplitz-quant-wo-measure-ip}.

\section{The Setting}
\label{Setting-section}

We will study an algebra $\mathcal{A}$ over the complex numbers $\mathbb{C}$ with unit $1$ 
and with an involutive, anti-linear {\em conjugation} (also called a {\em $*$-operation}), denoted by $f^*$ for
$f \in \mathcal{A}$, together with
a unital sub-algebra $\mathcal{P}$ of $\mathcal{A}$ (that is, with $1 \in \mathcal{P}$). 
We assume that $1^* = 1$. 
The algebra $\mathcal{A}$ will be the space of symbols for the Toeplitz operators which
we will define later.
So $\mathcal{A}$ is the `classical' space we wish to quantize.
For example, it could be an algebra of `functions' on a non-commutative phase space,
but in this paper we do not impose a Poisson structure on $\mathcal{A}$. 
The typical case that we have in mind is that $\mathcal{P}$ is not closed under the conjugation and that
in fact the intersection $\mathcal{P} \cap \mathcal{P}^*$ is as small as it possibly could
be, namely $\mathbb{C} 1$. 
Here $\mathcal{P}^* := \{ g^* \, | \, g \in \mathcal{P} \}$.
However, we will make no hypothesis about $\mathcal{P} \cap \mathcal{P}^*$. 

We do {\em not} assume that $\mathcal{A}$ is a $*$-algebra, namely that $(fg)^* = g^* f^*$ holds
for all $f,g \in \mathcal{A}$.
Also, we do not put any restriction on the dimensions of these vector spaces.
The existence of a unit in $\mathcal{A}$ is not an essential element of this theory, and many
of the results go through without assuming that it exists.
The fact that we are using an algebra 
of symbols allows us to include non-commutative geometry as a special case of this theory.
So $\mathcal{A}$ could be any algebra that is considered to be a non-commutative space.
The possibility of such a theory was raised, but not realized, in a remark in \cite{ali-englis2}.
 
We suppose there is a sesquilinear, complex symmetric form (or inner product)
 $\mathcal{A} \times \mathcal{A} \to \mathbb{C}$,
which is denoted by $\langle \cdot , \cdot \rangle_{\mathcal{A}}$.
Our convention is that this form is anti-linear in the first entry and linear in the second.
We allow the possibility that this inner product could be degenerate.
However, we impose the requirement that, when this inner product is restricted to
$\mathcal{P}$, it is positive definite.
Therefore $\mathcal{P}$ is a pre-Hilbert space.
We let $\mathcal{H}$ denote the completion of $\mathcal{P}$.
Therefore $\mathcal{P}$ can be realized as a dense subspace in the Hilbert space $\mathcal{H}$
with no loss of generality.
We assume from now on that this is the case. 
The letter $\mathcal{P}$ is meant to remind us of `polynomial' and `pre-Hilbert space'. 

We suppose there exists an orthonormal
indexed set $\Phi = \{ \varphi_j \, | \, j \in J \}  \subset \mathcal{P}$ that satisfies the following three conditions: 
\begin{enumerate}

\item $\Phi$ is a Hamel basis of $\mathcal{P}$.
 (A Hamel basis of a vector space is a maximal, linearly independent subset of 
 that vector space.)
 So, 
 $$
 \mathcal{P} = \Big\{ \sum_k a_k \varphi_k \, \Big| \, a_k = 0 \mathrm{~for~all~but~finitely~many~} k \Big\}.
$$

\item $\Phi$ is an orthonormal basis of $\mathcal{H}$.
So, 
$$
\mathcal{H} = \Big\{ \sum_k a_k \varphi_k \, \Big| \, \sum_k | a_k|^2 < \infty \Big\}.
$$
This is actually a consequence of Condition~1, but for the sake of clarity we state it
separately. 

\item For every $f \in \mathcal{A}$, the set defined by 
$\Phi_f := \{ \varphi_j \in \Phi \, | \, \langle \varphi_j, f \rangle_{\mathcal{A}} \ne 0 \}$
is finite. However, the cardinality of the set $\Phi_f$ can depend on $f$.

\end{enumerate}

If $\mathcal{P}$ is finite dimensional, as in \cite{part1}, then $\mathcal{P} = \mathcal{H}$
and such a subset $\Phi$ exists.
Notice that neither the $*$-operation nor the unit $1$ is mentioned 
in these three conditions and so some of
this theory can be developed without those structures, though the more interesting results
do use those structures. 
(Cp. Theorem~\ref{T33}, Parts~1, 4 and 5.)
We now fix such a set $\Phi$ and continue developing this theory further.
We address later the question if the subsequent theory really
depends on the choice of $\Phi$.
These conditions are technical in nature and could be modified to give similar theories. 
The first two conditions concern relations among $\Phi$, the pre-Hilbert space $\mathcal{P}$ 
and its completion $\mathcal{H}$.

The third condition has a quite different character, since it relates the sub-algebra $\mathcal{P}$ with
the larger algebra $\mathcal{A}$. 
(Recall $\mathcal{A}$ will be the symbol space.) 
Intuitively, this condition says that $\mathcal{A}$ is not `too' big or equivalently 
that $\mathcal{P}$ is not `too' small.
In particular, using Condition~3
the Toeplitz operators which we will define presently leave their common domain
invariant and so the composition of two of them is completely straightforward.
An alternative condition, weaker than Condition~3, would be to require: 
\begin{center}
$3^\prime.$ The series $\sum_j | \langle   \varphi_j , f \rangle_{ \mathcal{A} } |^2$ 
converges for all $f \in \mathcal{A}$.
\end{center}
With this weaker condition the common domain of our Toeplitz operators need not
be invariant and so the composition of them becomes problematic.
For example, the discussion of the canonical 
commutation relations becomes more complicated in this case.

The intuition behind this theory comes from physics. 
There are three spaces here, each with a physical interpretation: 
$\mathcal{A}$, $\mathcal{P}$ and $\mathcal{H}$. 
The starting point is $\mathcal{A}$, which corresponds to the `functions' on the classical phase space
of a classical physical system. 
The classical phase space is typically the space of positions and (linear) momenta 
of the physical system under consideration. 
However, here we allow a certain level of `quantum'  (i.e., non-commutative) behavior even at 
this `classical' level, since $\mathcal{A}$ need not be commutative. 
But we consider a fully quantum description of a physical system to have
at least two essential aspects: 
densely defined linear operators acting in a complex Hilbert space 
and Planck's constant $\hbar > 0$. 
The second space of the theory is the sub-algebra $\mathcal{P}$ of $\mathcal{A}$. 
The idea is that $\mathcal{P}$ corresponds to the functions on the configuration space 
of the physical system. 
The configuration space is typically the space of positions, and so has half the number of
variables as the phase space. 
Here $\mathcal{P}$ corresponds to the holomorphic polynomials (say) on the configuration space. 
With that interpretation the conjugate space $\mathcal{P}^*$ corresponds to the 
anti-holomorphic polynomials on the configuration space. 
These interpretations of $\mathcal{P}$ and $\mathcal{P}^*$ 
can be interchanged with no loss of mathematical 
generality nor of physical intuition. 
However, the central role of the conjugation operation $^*$ must be heavily emphasized. 
The third space of this theory is the Hilbert space $\mathcal{H}$, where the densely defined 
Toeplitz operators act. 
The quantum side of the theory resides in $\mathcal{H}$ and its Toeplitz operators. 
Each Toeplitz operator $T_g$ comes from an element $g \in \mathcal{A}$ called its symbol. 
The collection of all the Toeplitz operators gives us the Toeplitz quantization of $\mathcal{A}$. 
As a further step in the theory, Planck's constant is introduced into the quantum side of this
theory via the canonical commutation relations that certain Toeplitz operators (those of 
creation and annihilation) satisfy. 
The bridge between the classical side of the theory and the quantum side is provided 
by $\mathcal{P}$. 
Two algebraic structures, namely the multiplication and inner product for $\mathcal{A}$, 
are auxiliary to this basic outline and can be modified without changing the basic theory 
as indicated in \cite{toeplitz-quant-wo-measure-ip}. 

However, a quantization does not necessarily involve dynamics, that is, an equation of motion. 
It is rather well understood and accepted that the dynamics 
must be introduced in various ways, each way corresponding to 
its own quantum physical system. 
This is typically done by 
choosing a specific quantum Hamiltonian operator 
for the quantum system under consideration and using it in Schr\"odinger's equation. 
While the quantum Hamiltonian can be the quantization of a classical physics Hamiltonian, 
this aspect of quantization will not be considered in this paper beyond one simple remark for now.
One can always take a self-adjoint symbol $g$, that is $g^* = g$, and consider it as a
`classical' Hamiltonian whose corresponding quantum Hamiltonian is some self-adjoint 
extension of the symmetric Toeplitz operator $T_g$. 
Note that $T_g$ will be a symmetric operator under a hypothesis relating the inner product
with the conjugation. 
But in some examples that hypothesis need not hold and so the construction of self-adjoint
Hamiltonians becomes an important problem; 
one such example is given by $SU_q(2)$. 
We will return to this point. 

This paragraph is a non-rigorous discussion which is only meant to serve as motivation.
We first consider the formal sum
\begin{equation}
\label{formal-sum}
        K:= \sum_{j \in J} \varphi_j^* \otimes \varphi_j.
\end{equation}
We emphasize that the cardinality of the index set $J$ is completely arbitrary. 
If we restrict $j$ in the previous sum to lie in some finite subset of $J$, this gives a well-defined
element in $\mathcal{P}^* \otimes \mathcal{P}$, which in turn can be identified (essentially
by thinking of Dirac's bra-ket notation) as a finite rank projection operator mapping $\mathcal{P}$
to itself.
If $\mathcal{H}$ has finite dimension, then (\ref{formal-sum}) itself immediately identifies $K$ as the kernel of the
identity operator of $\mathcal{P} = \mathcal{H}$.
If $\mathcal{H}$ has infinite dimension, then (\ref{formal-sum}) also identifies $K$ as the kernel of the identity
operator of $\mathcal{P} $ provided that we interpret the infinite sum in the topology 
corresponding to the strong operator topology of bounded operators. 
So (\ref{formal-sum}) is basically a resolution of the identity of  $\mathcal{P}$.
It seems reasonable to suppose that (\ref{formal-sum}) could be replaced with a resolution
of the identity of $\mathcal{P}$ by coherent states without changing this theory dramatically.
However, we must emphasize that the Toeplitz quantization to be defined below is not the
coherent state quantization (see \cite{gazeau}) associated to (\ref{formal-sum}).
The latter quantization in this setting maps the function $\alpha : J \to \mathbb{C}$ to the operator
associated to $\sum_{j \in J} \alpha(j) \varphi_j^* \otimes \varphi_j$, modulo the usual technical details 
about convergence of the sum.
Moreover, the set of all such $\alpha$'s forms a commutative algebra, while we will quantize the
possibly non-commutative algebra $\mathcal{A}$.

A formal computation now gives for all $f \in \mathcal{P}$ that
\begin{equation*}
\langle K, f \rangle_{ \mathcal{A} } = \langle \sum_j \varphi_j^* \otimes \varphi_j , f \rangle_{ \mathcal{A} }
= \sum_j \langle  \varphi_j^* \otimes \varphi_j , f \rangle_{ \mathcal{A} }
= \sum_j \langle   \varphi_j , f \rangle_{ \mathcal{A} } \,\, \varphi_j.
\end{equation*}
These remarks motivate this \textit{formal} definition  for all $f \in \mathcal{P}$:
\begin{equation*}
\langle K, f \rangle_{ \mathcal{A} } 
:= \sum_j \langle   \varphi_j , f \rangle_{ \mathcal{A} } \,\, \varphi_j.
\end{equation*}
Even though the conjugation was used to motivate this definition, 
note that the conjugation does not
appear in the definition.
By Condition~3 the sum on the right side has only finitely many non-zero terms.
It gives us an element in $\mathcal{P}$, since each $\varphi_j \in \mathcal{P}$.
Moreover, since $f \in \mathcal{P}$ we have that 
$$
    \sum_j \langle   \varphi_j , f \rangle_{ \mathcal{A} } \,\, \varphi_j = f,
$$
which is an elementary result.
So $\langle K, f \rangle_{ \mathcal{A} } = f$ and thus
$K$ can also be viewed formally as a generalized reproducing kernel for $\mathcal{P}$.
While all the material of this paragraph can be developed rigorously
(for example, following the presentations in \cite{part1} or \cite{quantum-plane}), 
for the moment we merely wished to give an idea of what the set $\Phi$ is good for.

It is natural to require that the inner product in $\mathcal{A}$ has this relation with the
conjugation in $\mathcal{A}$:
\begin{equation}
\label{star-compatibility}
\langle f , g    \rangle_{ \mathcal{A} } ^* = \langle f^* , g^*    \rangle_{ \mathcal{A} }
\end{equation}
for all $f, g \in \mathcal{A}$.
This simply means that $f \mapsto f^*$ is an anti-unitary map of $\mathcal{A}$ to itself.
This condition is satisfied by the paragrassmann algebras (see~\cite{part1})
and by the complex quantum plane (see~\cite{quantum-plane}).
Here is an immediate consequence of this requirement.
\begin{prop}
$\mathcal{P}^*$ is a pre-Hilbert space with respect to the restriction of the inner product
$\langle \cdot , \cdot \rangle_{\mathcal{A}}$ to it. 
\end{prop}
\textbf{Sketch of Proof:} 
The set $\Phi^* = \{ \varphi_j^* \, | \, j \in J \}$ is an orthonormal set in $\mathcal{P}^*$
by (\ref{star-compatibility}).
It is left to the reader to prove that $\Phi^*$ is a Hamel basis of $\mathcal{P}^*$ as well.
Then it follows that the inner product $\langle \cdot , \cdot \rangle_{\mathcal{A}}$
restricted to $\mathcal{P}^*$ is positive definite, and so $\mathcal{P}^*$ is a pre-Hilbert space.
$\quad \blacksquare$

\vskip 0.4cm 
We denote the completion of the pre-Hilbert space $\mathcal{P}^*$ by $\mathcal{H}^*$.
There is an anti-unitary identification \textit{as inner product spaces} 
between the pair of spaces $(\mathcal{P}, \mathcal{H})$ and the
pair of spaces $(\mathcal{P}^*, \mathcal{H}^*)$ induced by 
$\mathcal{P} \ni f \mapsto f^* \in \mathcal{P}^*$.
We sometimes refer to $\mathcal{H}$ as the holomorphic space (or Segal-Bargmann space)
and to $\mathcal{H}^*$ as the anti-holomorphic space (or anti-Segal-Bargmann space).
It turns out that these designations are completely arbitrary and can be reversed with absolutely
no loss of rigor nor (if one is savvy enough) of intuition.
Since $\mathcal{P} \cap \mathcal{P}^*$ consists either way of elements which, 
according to this classification, are both holomorphic
and anti-holomorphic, we see the intuition behind the condition $\mathcal{P} \cap \mathcal{P}^* = \mathbb{C} 1$.
But we continue with $\mathcal{P} \cap \mathcal{P}^*$ being completely arbitrary.

Curiously, the statement  ``$\mathcal{P}^*$ is a sub-algebra of $\mathcal{A}$'' 
could be false, though it is true whenever $\mathcal{A}$ is a $*$-algebra or for
the examples in \cite{part2} and \cite{quantum-plane} (even when $\mathcal{A}$ is not  a $*$-algebra). 
In \cite{part2} $\mathcal{P}$ is the holomorphic Hilbert space, while 
in \cite{quantum-plane} the sub-algebra $Pre(\theta)$ plays the role of $\mathcal{P}$.

\begin{prop}
If $\big( \mathcal{A}, \mathcal{P}, \Phi, \langle \cdot , \cdot \rangle_{ \mathcal{A} } \big)$ satisfy Conditions 1--3,
then it follows that
$\big( \mathcal{A}, \mathcal{P}^*, \Phi^*, \langle \cdot , \cdot \rangle_{ \mathcal{A} } \big)$ satisfy Conditions 1--3.
\end{prop}
\textbf{Sketch of Proof:} 
We have already commented that $\Phi^*$ satisfies Condition~1. 
And, as also noted, Condition~2 readily follows from Condition~1.
That Condition~3 holds we leave to the reader as a quick exercise using (\ref{star-compatibility}).
$\quad \blacksquare$

\vskip 0.4cm

In the rest of this paper we will only use this relation between the sesquilinear form and the
conjugation:
\begin{equation}
\label{form-conjugation-compatiblity} 
 \langle f_1 , f_2 g \rangle_{ \mathcal{A} } = \langle f_1 g^* , f_2  \rangle_{ \mathcal{A} }
 \qquad \mathrm{for~~} f_1, f_2 \in \mathcal{P}, \,\, g \in \mathcal{A}. 
\end{equation}
This identity holds for the examples in 
\cite{part2} and \cite{quantum-plane}.
So those two examples are special cases of the theory in the rest of this paper.
However, the example using $SU_q(2)$ in Section 
\ref{SUq2-section} does not satisfy this identity. 
Nonetheless, it is an illustrative example of many aspects of Toeplitz quantization.

\section{Toeplitz Quantization}
\label{TQ-section}

First we take any $g \in \mathcal{A}$ and use it to define
a linear map 
$$
       M_g : \mathcal{P} \to \mathcal{A}
$$
by $M_g \psi := \psi g $ for all $\psi \in \mathcal{P}$.
Notice that $\psi g \in   \mathcal{A}$, since it is a product of
two elements in the algebra $\mathcal{A}$.
In this paper this is the main use of the multiplication of $\mathcal{A}$.
So a bilinear map $\mathcal{P} \times \mathcal{A} \to \mathcal{A}$
could be used instead of the multiplication $(\psi, g) \mapsto \psi g$.
This map would have to satisfy some other conditions as well to make the theory work out.
We will not go into further details about this more general approach, which is discussed 
in \cite{toeplitz-quant-wo-measure-ip}.

We next wish to use the kernel $K$ to extend the identity map on $\mathcal{P}$
to a projection map $P_K : \mathcal{A} \to \mathcal{A}$.
The technique is standard in analysis.
We simply use the same {\em formula} to define a different {\em operator},
where the difference consists in using a different domain of definition.
So we define for $f \in \mathcal{A}$:
\begin{equation}
\label{define-PK}
   P_K f := \sum_j \langle   \varphi_j , f \rangle_{ \mathcal{A} } \,\, \varphi_j.
\end{equation}
Of course, by our previous discussion we have $P_K f = f$ provided that $f \in \mathcal{P}$.
Now for $f \in \mathcal{A}$ 
we have assumed that only finitely many of the coefficients 
$\langle   \varphi_j , f \rangle_{ \mathcal{A} }$ are non-zero.
So the  sum on the right side of (\ref{define-PK}) is effectively over a finite number of terms and
so $P_K f \in \mathcal{P}$ for all $f \in \mathcal{A}$, that is, $P_K : \mathcal{A} \to \mathcal{P}$.
It is important to emphasize that the inner product on $\mathcal{A}$ is used in this theory only 
to define this linear map $P_K$. 

\begin{theorem}
\label{T31}
$P_K$ is a projection, that is $P_K^2 = P_K$, and is  symmetric with respect to
the inner product on $\mathcal{A}$, that is
\begin{equation}
\label{PK-is-symmetric}
    \langle P_K f , g \rangle_{ \mathcal{A} } = \langle  f , P_K g \rangle_{ \mathcal{A} } 
\end{equation}
for all $f,g \in \mathcal{A}$.
If the inner product is non-degenerate, then we can write (\ref{PK-is-symmetric})
as $P_K^* = P_K$, where $P_K^*$ is {\em the unique} adjoint operator of $P_K$.
\end{theorem}
\textbf{Proof:} 
First we note that $P_K f \in \mathcal{P}$ for all $f \in \mathcal{A}$ and that $P_K$
acts as the identity on $\mathcal{P}$.
So $P_K (P_K f ) = P_K f $ for all $f \in \mathcal{A}$, thereby proving that $P_K^2 = P_K$.
Next, one readily calculates that each side of (\ref{PK-is-symmetric})
is equal to 
$\sum_j \langle f, \varphi_j \rangle_{ \mathcal{A} } \langle \varphi_j, g \rangle_{ \mathcal{A} } $.
And this is a sum with only finitely many non-zero terms, and so there is no problem with the convergence of
this sum.
$\quad \blacksquare$
\vskip 0.4cm \noindent

We now return to the question whether this theory depends on the choice
of orthonormal set $\Phi = \{ \varphi_j \, | \, j \in J \}$.
The point is that this set is used to define $P_K$. 
But suppose that $\Psi = \{ \psi_j \, | \, j \in J \}$ is another orthonormal set in 
$\mathcal{P}$ that is also a Hamel basis of $\mathcal{P}$.
And let $P_K$ be the projection operator defined above using the set $\Phi$.
We temporarily denote $P_K$ by $P_K^\Phi$ to indicate its dependence on $\Phi$.
Suppose that $f \in \mathcal{A}$.
Then as we have seen $P_K^\Phi f \in \mathcal{P}$.
So we can expand $P_K^\Phi f$ uniquely in the Hamel basis $\Psi$ of $\mathcal{P}$ to get
\begin{equation}
\label{expand-in-Psi}
    P_K^\Phi f = \sum_k a_k \psi_k 
\end{equation}
with all but finitely many $a_k =0$.
Taking the inner product of this with $\psi_j$ yields
\begin{equation}
\label{aj-equals}
      a_j = \langle \psi_j, P_K^\Phi f \rangle_{ \mathcal{A} } = \langle P_K^\Phi  \psi_k,  f \rangle_{ \mathcal{A} } 
            = \langle \psi_k,  f \rangle_{ \mathcal{A} } 
\end{equation}
for all $j \in J$, since $\Psi$ is orthonormal and $P_K^\Phi$ acts as the identity of $\mathcal{P}$.
So the set $\Psi_f := \{ \psi_j \, | \, \langle \psi_j, f \rangle_{\mathcal{A}} \ne 0 \}$
is finite. 

Substituting (\ref{aj-equals}) back into (\ref{expand-in-Psi}) we see for all $f \in \mathcal{A}$ that
\begin{equation*}
      P_K^\Phi f = 
      \sum_k \langle \psi_k,  f \rangle_{ \mathcal{A} } \, \psi_k
      = P_K^\Psi f, 
\end{equation*}
using in the second equality the corresponding definition of the projection operator 
$ P_K^\Psi$ defined by the set $\Psi$.
In short, the definition of $P_K^\Phi$ does not depend on the choice of the
set $\Phi$.
Since the only essential use of the set $\Phi$ is exactly to define $P_K^\Phi$,
we now have shown that the subsequent theory does not depend on the
particular choice $\Phi$.
And so we revert to our original notation: $P_K$.

Since $\Phi$ will make only minor appearances in the rest of this paper 
and it was only used so far 
to define $P_K$, an alternative approach to this theory is to start with $\mathcal{A}$,
$\mathcal{P}$ and an inner product on $\mathcal{P}$, all as above. 
But instead of $\Phi$ one introduces
an operator $P : \mathcal{A} \to \mathcal{P}$ straightaway 
with the properties in Theorem~\ref{T31}.

\begin{definition}
For any $g \in \mathcal{A}$ we can form the composition of linear maps
$$
 \mathcal{P} \stackrel{M_g}{\longrightarrow} \mathcal{A} \stackrel{P_K}{\longrightarrow} \mathcal{P} 
$$
which we define to be the {\em Toeplitz operator associated with the symbol $g \in \mathcal{A}$},
denoted by $T_g := P_K M_g$.
\end{definition}

Notice that $T_g$ is defined in the dense domain $Dom (T_g) := \mathcal{P}$, which does not
depend on $g$.
Furthermore, $\mathcal{P}$ is invariant under the action of $T_g$ and so we can always
compose any finite number of Toeplitz operators. 
This is not a usual situation in Toeplitz operator theory in function spaces, where the
domain typically depends on the symbol and where that domain is not necessarily invariant.

The symbol $g$ in $T_g$ is known as the upper symbol in Lieb's paper \cite{lieb} 
and as the  contravariant symbol in Berezin's paper \cite{berezin}. 
The corresponding lower or covariant symbol of those papers does not seem to have an
exact analogue in this general non-commutative setting.
For example, see \cite{csq} where lower symbols are introduced in a non-commutative
setting that includes coherent states. 
So it may well be a worthwhile avenue for future research
 to modify the present theory so as to include coherent states as well.

The linear map $T : g \mapsto T_g$ for $g \in \mathcal{A}$ is called the \textit{Toeplitz quantization}.
Since $\mathcal{A}$ can be a non-commutative algebra, we do include the
possibility that the symbols of the Toeplitz quantization do not commute among themselves.
This is in rather sharp contrast to most studies of Toeplitz operators in classical analysis, 
where the symbols are real or complex valued functions with multiplication defined pointwise. 
Other definitions in the literature of Toeplitz operators with non-commuting symbols 
will be discussed in the concluding section.  
This is a strictly mathematical point of view of what is being done here. 

However, from a physics point of view, we are quantizing the (possibly non-commutative, i.e., quantum) 
space $\mathcal{A}$ by densely defined operators acting 
in the quantum Hilbert space $\mathcal{H}$.
The space $\mathcal{A}$ could be the functions on a phase space or just about
anything else.
When $\mathcal{A}$ is non-commutative
this can be considered as a type of second quantization (that is, it is the quantization of something
that is already quantum), 
though the result of the quantization is not a quantum field theory by any means.
However with a little bit more work, this Toeplitz quantization can be realized as a functor and
so is in accord with Nelson's maxim that second quantization is a functor.
(See Section X.7 in \cite{reed-simon2}.)

We can take the co-domain of the Toeplitz quantization $T$ to be the vector space
$$
\mathcal{L} \equiv
 \mathcal{L} (\mathcal{P}) := \{ S: \mathcal{P} \to \mathcal{P} \, | \, S \mathrm{~is~linear} \}
$$
of densely defined linear operators in $\mathcal{H}$ with common invariant domain $\mathcal{P}$.
So, $\mathcal{L} (\mathcal{P})$ is an algebra under composition, though it does
not have a natural norm, thereby putting it on the same footing as the algebra $\mathcal{A}$.
Then the Toeplitz quantization
$$
          T: \mathcal{A} \to \mathcal{L} (\mathcal{P}) 
$$
is a linear map between algebras. 
However, it is not expected to be an algebra morphism in any reasonable set-up.
Nonetheless, it does have some properties related to the multiplication as well as some 
other nice properties.
Parts~2 and 3 below can be false without the hypothesis that
$\mathcal{P}$ is a sub-algebra.

\begin{theorem}
\label{T33}
The following hold:
\begin{enumerate}

\item $T_1 = I_{\mathcal{P}}$ where $I_{\mathcal{P}}$ is the identity map of $\mathcal{P}$.

\item If $g \in \mathcal{P}$, then $T_g = M_g$.

\item If $g \in \mathcal{A}$ and $h \in \mathcal{P}$, then $T_{g} T_{h} = T_{h g} $.

\item Suppose that $f_1, f_2 \in \mathcal{P}$ and $g \in \mathcal{A}$.
Then
\begin{equation}
\label{adjoint-relation}
\langle T_g f_1 , f_2 \rangle_{ \mathcal{A} } = \langle  f_1 , T_{g^*} f_2 \rangle_{ \mathcal{A} }.
\end{equation}
This can also be expressed by saying that $T_{g^*} \subset (T_g)^*$ or equivalently that
$T_{g} \subset ( T_{g^*} )^*$.
This shows a compatibility between the $*$-operation in the algebra $\mathcal{A}$ and the adjoint
operation of densely defined operators.

\item If $g = g^*$, then $T_g$ is a symmetric operator.

\end{enumerate}
\end{theorem}
\textbf{Proof:} 
For Part~1, we note that $M_1$ is just the inclusion map of $\mathcal{P}$
into $\mathcal{A}$. Since $P_K$ acts as the identity on $\mathcal{P}$,
we get $T_1 = P_K M_1 = I_{\mathcal{P}}$.

For Part~2 we remark that the range of $M_g$ is contained in $\mathcal{P}$
for $g \in \mathcal{P}$, since $M_g \psi = \psi g \in \mathcal{P}$ for all 
$\psi \in \mathcal{P}$.
Here we are using the hypothesis that $\mathcal{P}$ is a sub-algebra of $\mathcal{A}$.
But $P_K$ acts as the identity on $\mathcal{P}$. 
So, $T_g = P_K M_g = M_g$.

For Part~3 we first note for all $g \in \mathcal{A}$ and all $h, \phi \in \mathcal{P}$ that
$$
       M_g M_h \phi = (M_h \phi ) g = (\phi h ) g = \phi (h  g)= M_{h g} \phi.
$$
So, using this and Part~2 we see that
$$
     T_{g} T_{h} 
     = P_K M_g M_h = P_K M_{h g} = T_{h g}. 
$$

For Part~4 we suppose $f_1, f_2 \in \mathcal{P}$ and $g \in \mathcal{A}$. 
Then we calculate 
\begin{align*}
\langle T_g f_1, f_2 \rangle_{ \mathcal{A} } &= \langle P_K M_g f_1, f_2 \rangle_{ \mathcal{A} } 
= \langle f_1 g, P_K f_2 \rangle_{ \mathcal{A} }
= \langle f_1 g,  f_2 \rangle_{ \mathcal{A} },
\end{align*}
where the last equality follows from $f_2 \in \mathcal{P}$.
Similarly we see that
\begin{align*}
\langle  f_1, T_{g^*} f_2 \rangle_{ \mathcal{A} } &= \langle  f_1, P_K M_{g^*} f_2 \rangle_{ \mathcal{A} } 
 = \langle   P_K f_1,  f_2 g^* \rangle_{ \mathcal{A} }
 = \langle f_1,  f_2 g^* \rangle_{ \mathcal{A} },
\end{align*}
where now the last equality follows from $f_1 \in \mathcal{P}$.
Finally, we use the identity 
$\langle f_1,  f_2 g^* \rangle_{ \mathcal{A} } = \langle f_1 g, f_2 \rangle_{ \mathcal{A} }$,
which we took as a hypothesis in (\ref{form-conjugation-compatiblity}). 

Next, the two relations $T_{g^*} \subset (T_g)^*$ and
$T_{g} \subset ( T_{g^*} )^*$ follow immediately from (\ref{adjoint-relation}).
These relations are equivalent using the substitution $g \mapsto g^*$.

Part~5 is an immediate consequence of Part~4 and the definition (see \cite{reed-simon1}) of a
symmetric operator.
$\quad \blacksquare$

\vskip 0.4cm 
We now are presented with a classical problem in the functional analysis
of densely defined operators, namely, in the case of a  self-adjoint symbol $g = g^*$ we have
the symmetric, densely defined operator $T_g$.
One would like to know whether this operator has self-adjoint
extensions and, if it does, how to explicitly classify them.
In particular, it could be that $T_g$ is self-adjoint or essentially
self-adjoint for particular choices of self-adjoint $g$.
For example, $T_1 = I_{ \mathcal{P} }$ is essentially self-adjoint.
Of course, the probabilistic interpretation of any self-adjoint extension of 
$T_g$ as a physical observable 
would be based on its projection valued measure, just as
is done in \cite{reed-simon1}.

We continue with some technical, but important, mathematical details.
\begin{theorem}
For any $g \in \mathcal{P}$, the Toeplitz operator $T_g$ is closable
and its closure $\overline{T_g}$ satisfies
$$
     \overline{T_g} = (T_g)^{**} \subset ( T_{g^*} )^*.
$$
\end{theorem}
\textbf{Proof:} All of this follows from basic functional analysis. (See \cite{reed-simon1}.)
For example, an operator $R$ is closable if and only if $Dom \, R^*$ is dense.
But $Dom (T_g)^* \supset Dom \, T_{g^*} = \mathcal{P}$ and $\mathcal{P}$ is
dense in $\mathcal{H}$. 
So, $Dom (T_g)^*$ is also dense and therefore $T_g$ is closable. 
Next $ \overline{T_g} = (T_g)^{**}$ comes directly from \cite{reed-simon1}.
Finally, $T_{g^*} \subset (T_g)^*$ implies that $ (T_g)^{**} \subset ( T_{g^*} )^*$.
$\quad \blacksquare$

\vskip 0.4cm \noindent
Since $T_g$ is closable, we would expect that in this setting there is a more explicit description 
of its closure $\overline{T_g} = (T_g)^{**} $.
We leave this as a problem for future consideration.

Many other problems that are considered in the usual, classical Toeplitz quantization of functions
also arise in this non-commutative context. 
These include finding necessary conditions as well as sufficient conditions for a Toeplitz operator to
be bounded.
Then given that a Toeplitz operator is bounded, there are open problems remaining to find 
 necessary conditions as well as sufficient conditions for it to be compact, to be in a Schatten class,
to be normal, to be unitary and so forth.
However, these questions are known to depend on the particular properties of $\mathcal{P}$ and $\mathcal{A}$
in the case of classical Toeplitz operators and so may not be amenable to much more analysis in 
this general setting.

The material in this section deals with one of eight (at least!) possible Toeplitz quantizations
that can be defined in this setting. 
For starters, one could change the definition of the 
operators $M_g$ to be multiplication on the left (instead of on the right) by $g$.
This would give us a different, but very similar theory. 
Another possibility is to consider the Toeplitz quantization given by Toeplitz operators acting in
the anti-holomorphic space $\mathcal{H}^*$ together with the two options for how the
multiplication operators $M_g$ act, namely, on the right or on the left.
This gives us two more Toeplitz quantizations \textit{provided that} $\mathcal{P}^*$ is a sub-algebra of $\mathcal{A}$.
Again, these are quite similar to the theory developed here.
And yet another variation is to replace $M_g$ with $M_{g^*}$ in each of the previous four cases,
thereby resulting in anti-linear Toeplitz quantizations. 
But these are all minor variations on the same theme and will not be
discussed further.

\section{Seeking $\mathcal{P}$}
\label{section-seeking-P}

In practice, we usually have a candidate algebra $\mathcal{A}$ which we wish to quantize. 
The `tricky bit' is to find the appropriate sub-algebra $\mathcal{P}$ (as well as the inner
product, of course) in order to make everything work out. 
In this section we consider various aspects of this situation. 
We start off with some elementary results. 

\begin{lemma}
Suppose that $\mathcal{H}$ is a separable Hilbert space and 
that $\mathcal{P} \subset \mathcal{H}$ is a dense subset of $\mathcal{H}$. 
Then there exists $D^\prime \subset \mathcal{P}$ such that $D^\prime$ is a countable dense 
subset of $\mathcal{H}$, and hence also a countable dense subset of $\mathcal{P}$. 
\end{lemma}
\textbf{Remark:} 
The notation betrays how we will use this result. 
In particular, we will be interested in the case when $\mathcal{P}$ is a dense 
\textit{subspace} of $\mathcal{H}$. 
In that case $\mathcal{P}$ will not be countable, except in the trivial case 
when $\mathcal{H} = 0$. 

\vskip 0.4cm \noindent 
\textbf{Proof:} 
Let $D \subset \mathcal{H}$ be a dense countable subset of $\mathcal{H}$. 
For each $x \in D$ and each integer $n \ge 1$ there exists $y = y(x,n) \in \mathcal{P}$ 
such that 
$$
       || x - y || = || x - y(x,n) || < 1/n,
$$
since $\mathcal{P}$ is dense in $\mathcal{H}$. 
Define
$$
     D^\prime := \cup_{ (x,n) } \, \{ y(x,n) \},
$$
where $(x,n) \in D \times \mathbb{N}^+ $, a countable set. 
So $D^\prime$ is countable. 
Obviously by construction $D^\prime \subset \mathcal{P}$. 

To show that $D^\prime$ is dense in $\mathcal{H}$ we take $z \in \mathcal{H}$ arbitrary as well 
as $\epsilon > 0$ arbitrary. 
Pick an integer $n$ sufficiently large so that 
$$
       \dfrac{1}{n} < \dfrac{\epsilon}{2}. 
$$
Then there exists $x_0 \in D$ such that $|| z - x_0 || < \epsilon / 2$, since 
$D$ is dense in $\mathcal{H}$. 
Then,
$$
  || z - y (x_0,n) || \le ||z - x_0 || + || x_0 - y (x_0,n) || < \dfrac{\epsilon}{2} + \dfrac{1}{n} < \epsilon. 
$$
But $y (x_0,n) \in D^\prime$. 
So $D^\prime$ is dense in $\mathcal{H}$. 
Even more so, this means that $D^\prime$ is also dense in $\mathcal{P}$. 
$\quad \blacksquare$

\begin{prop}
Suppose that $\mathcal{Q} \subset \mathcal{H}$ is a dense subspace of a 
separable Hilbert space $\mathcal{H}$. 
Then there exists an orthonormal basis $\{ \varphi_j ~|~ j \ge 0 \}$ of $\mathcal{H}$ 
such that $\varphi_j \in \mathcal{Q}$ for all $j \ge 0$. 
\end{prop}
\textbf{Proof:} 
If $\mathcal{H}$ is finite dimensional, then $\mathcal{Q} = \mathcal{H}$ and so 
the result is trivial.
So from now on we assume that $\mathcal{H}$ is infinite dimensional.

By the previous lemma there exists some (highly non-unique) 
$D^\prime \subset \mathcal{Q}$ such that 
$D^\prime$ is a dense countable subset of $\mathcal{H}$. 
Use some (highly non-unique) bijection of $D^\prime$ with $\mathbb{N}$ to express $D^\prime$ 
as a sequence. 
Applying Gram-Schmidt to this sequence we obtain a countable orthonormal basis 
$\{ \varphi_j ~|~ j \ge 0 \}$ of $\mathcal{H}$. 
Now Gram-Schmidt produces the elements $\varphi_j$ as finite linear combinations of 
elements in $D^\prime$. 
But $D^\prime$ lies inside the subspace $\mathcal{Q}$. 
So each $\varphi_j \in \mathcal{Q}$. 
$\quad \blacksquare$

\vskip 0.4cm \noindent 
In general, the orthonormal basis $\{ \varphi_j \}$ will not be a Hamel basis 
of $\mathcal{Q}$. 
Simply stated, $\mathcal{Q}$ could be too big. 
For example, $\mathcal{Q} = \mathcal{H}$ is a possibility, in which case it 
is well known that $\{ \varphi_j \}$ is not a Hamel basis of $\mathcal{Q}$.
However, we can define
$$
     \mathcal{P}:= \mathrm{span} \{ \varphi_j \},
$$
the algebraic span of the orthonormal basis. 
Then it is clear that 
$\{ \varphi_j \}$ is a Hamel basis of $\mathcal{P}$.
This procedure of going from a dense subspace 
$\mathcal{Q}$ of $\mathcal{H}$ to a `good' dense subspace 
$\mathcal{P}$ of $\mathcal{H}$ is not an algorithm. 
Many such subspaces $\mathcal{P}$ will be produced in general. 

So we would like some criteria for which of these `good' subspaces 
$\mathcal{P}$ are `better' than others. 
And also we would like some concept of what constitutes the `best' 
such subspace $\mathcal{P}$. 
Here is how one can go about doing that. 
But now the emphasis is not on the relation of $\mathcal{P}$ with the
quantum Hilbert space $\mathcal{H}$, but rather how to look for an
adequate sub-algebra $\mathcal{P}$ inside of a given algebra $\mathcal{A}$. 

To address this situation let us first suppose that at least one such  
 $\mathcal{P}$ has been found. 
 Say that $\mathcal{P}^\prime$ is a sub-algebra of $\mathcal{P}$.
 Since the inner product on $\mathcal{A}$ is positive definite when restricted to $\mathcal{P}$, 
 it is also positive definite when restricted to $\mathcal{P}^\prime$. 
 Consequently, $\mathcal{P}^\prime$ is a pre-Hilbert space, and so its completion, denoted as 
 $\mathcal{H}^\prime$, is a closed subspace of $\mathcal{H}$. 
 Suppose that we can find an orthonormal set $\Phi \subset \mathcal{P}$ 
 satisfying Conditions~1 to 3 and such that there exists
 a subset $\Phi^\prime$ of $\Phi$ which is a Hamel basis of $\mathcal{P}^\prime$. 
 Then the pair $(\mathcal{P}^\prime, \Phi^\prime)$ satisfies
 all the Conditions~1 to 3 for developing this theory. 
 The collection of all such sub-algebras $\mathcal{P}^\prime$ clearly form a partially ordered  
 system, where the partial order is the inclusion of one sub-algebra in another as well as the inclusion
of their Hamel bases.
This discussion serves as motivation for the 
next definition.  
 
\begin{definition}
Suppose $\mathcal{A}$ is an algebra with an inner product.
Let $(\mathcal{P}, \Phi)$ and $(\mathcal{P}^\prime, \Phi^\prime)$ satisfy the 
Conditions~1 to 3 with respect to the algebra $\mathcal{A}$ and the inner product of
$\mathcal{A}$ is positive definite on $\mathcal{P}$ and on $\mathcal{P}^\prime$. 
Then we say that $(\mathcal{P}^\prime, \Phi^\prime)$ is {\rm smaller than} 
$(\mathcal{P}, \Phi)$ provided that $\mathcal{P}^\prime \subset \mathcal{P}$ 
and $\Phi^\prime \subset \Phi$. 
\\
\textbf{Notation:} $(\mathcal{P}^\prime, \Phi^\prime) << (\mathcal{P}, \Phi)$. 
\end{definition}

It should be clear that $<<$ is a partial order. 
We are really interested in finding a maximal 
pair $(\mathcal{P}, \Phi)$ for a given algebra $\mathcal{A}$ with an inner product. 
Of course, the pair $(\mathcal{P}, \Phi) = (0, \emptyset)$ always satisfies the 
Conditions~1 to 3. 
This is the minimal pair. 
And it is trivial. 

\begin{prop}
Suppose $\mathcal{A}$ is an algebra with an inner product. 
Then there exists a maximal pair $(\mathcal{P}, \Phi)$ for $\mathcal{A}$ 
with respect to the partial order $<<$ 
such that Conditions~1 and 2 hold and the inner product of
$\mathcal{A}$ is positive definite on $\mathcal{P}$.  
\end{prop}
\textbf{Proof:} 
This is a simple application of Zorn's lemma. 
The point is that if $(\mathcal{P}_\alpha, \Phi_\alpha)$ is any ascending chain of pairs 
with respect to $<<$, 
then the pair $(\cup_\alpha \mathcal{P}_\alpha,  \cup_\alpha \Phi_\alpha)$ 
satisfies Conditions~1 and 2, but
not necessarily Condition~3. 
Also, the inner product of $\mathcal{A}$ is positive definitive when restricted to 
$\cup_\alpha \mathcal{P}_\alpha$, 
since it is positive definite on each $\mathcal{P}_\alpha$.
$\quad \blacksquare$

\vskip 0.4cm 
In practice, one usually wants the quantum Hilbert space to be separable. 
However, the maximal pair given in the previous proposition need not have 
$\Phi$ countable. 
The problem of finding a maximal pair satisfying 
Conditions~1 to 3 remains to be considered case by case, although the
considerations in this section could prove helpful. 
A further restriction which should simplify this problem is to require pairs for which
$\mathcal{P} \cap \mathcal{P}^* = \mathbb{C} 1$ holds.

\section{Creation and Annihilation Operators}
\label{CA-section}

\begin{definition}
Let $g \in \mathcal{P}$ be given.
Then the {\em creation operator} associated to $g$ is defined to be
$$
            A^\dagger (g) := T_g
$$
and the  {\em annihilation operator} associated to $g$ is defined to be
$$
            A (g) := T_{g^*}.
$$
\end{definition}
These are reasonable definitions given that they are in accord with
the usual meaning of these terms as exemplified in \cite{brian} and \cite{quantum-plane}.
However, there are other normalizations used as well for these operators.
One of these entails putting a factor of $\hbar^{-1/2}$ on the right sides of
these definitions, where $\hbar$ denotes Planck's constant.
But we will postpone the introduction of Planck's constant to a bit later.
Notice that $g \mapsto A^\dagger (g) $ is linear (as already remarked) and that
$g \mapsto A (g)$ is anti-linear.
Also $ A^\dagger (g) = T_g = M_g$ holds, since $g \in \mathcal{P}$.
Since $A^\dagger (1) = A(1) = T_1 = I_{ \mathcal{P} }$, we see that $I_{ \mathcal{P} }$
is both a creation and an annihilation operator. 
In fact for any $g \in \mathcal{P} \cap \mathcal{P}^*$, one has $T_g = A^\dagger (g) = A (g^*)$ and so
$T_g$ is both a creation and an annihilation operator. 

One of the important contributions of Bargmann's seminal paper \cite{bargmann} 
is that it realizes the creation and annihilation operators introduced by Fock as adjoints
of each other with respect to the inner product on the Hilbert space
which is nowadays called the Segal-Bargmann space. 
The creation operator $A^\dagger (g)$ and the annihilation operator $A (g)$
also have this relation, modulo domain considerations, as we have 
already seen in Theorem \ref{T33}, Part~4.
Whether each is \textit{exactly} the adjoint of the other as in \cite{bargmann} 
is an open question if $\mathcal{P}$ has infinite dimension, but is trivially so
for finite dimensional $\mathcal{P}$.

\section{Anti-Wick Quantizations}
\label{AW-quantization-section}

We now have the language needed to discuss whether this is an anti-Wick quantization,
as is expected from a Toeplitz quantization.
First recall that we have shown
\begin{equation}
\label{sometimes}
 T_{g h} = T_{h} T_{g}  
\end{equation}
provided that $g \in \mathcal{P}$ but with $h \in \mathcal{A}$ being arbitrary.
Because we are allowing non-commutative algebras $\mathcal{A}$,
we are led to two definitions for `anti-Wick'  in this theory.
These are clearly equivalent conditions if $\mathcal{A}$ is commutative as the reader will soon appreciate.
\begin{definition}
We say that $T$ is an  {\em  anti-Wick quantization} if 
$$
T_{h g^*} = T_{g^*} T_{h} 
$$
for all $g, h \in \mathcal{P}$.
We say that $T$ is an {\em alternative anti-Wick quantization} if 
$$
T_{g^* h} = T_{g^*} T_{h}  
$$
for all $g, h \in \mathcal{P}$.

\end{definition}
Notice that on the right side in both of these definitions we have the product of an
annihilation operator $T_{g^*} $ to the left of a creation operator $T_h$.
And so the right side is in anti-Wick order for each of these definitions.
The naming of these two properties was determined only after proving the following results. 
What we deem to call the anti-Wick quantization turns out to be the `correct' generalization
of this notion to the present setting as the next result shows.

\begin{theorem}
The Toeplitz quantization $T$ is an anti-Wick quantization.
\end{theorem}
\textbf{Proof:}
Take $g, h \in \mathcal{P}$.
Then 
$
    T_{h g^*} =  T_{g^*}  T_{h}
$,
where we have used (\ref{sometimes}).
$\quad \blacksquare$

\vskip 0.4cm \noindent
This clarifies why the examples in \cite{part2} and \cite{quantum-plane} 
are anti-Wick quantizations even though they arise in a
non-commutative context.
The longer, explicit computations given in those references are not
needed as we can now see.

This theorem has several immediate consequences:
\begin{corollary}
If $\mathcal{A} = \mathcal{P} \mathcal{P}^*$, then 
one can write any Toeplitz operator as a sum of terms in anti-Wick order.
\end{corollary}

\begin{corollary}
If $\mathcal{A}$ is commutative, then the Toeplitz quantization $T$ is an alternative
anti-Wick quantization.
\end{corollary}
The examples in \cite{part2} and \cite{quantum-plane} for $q \ne 1$ (which is the
non-commutative case) are \textit{not} alternative anti-Wick quantizations.

The last corollary has a partial converse.
\begin{theorem}
$T$ is not an alternative anti-Wick quantization if and only if 
there exist elements $g, h \in \mathcal{P}$ such that
$T_{g^* h } \ne T_{h g^*} $.
\end{theorem}
\textbf{Proof:}
As already shown  $T_{g^*}  T_{h} = T_{h g^*}$ is an identity for all $g, h \in \mathcal{P}$.
Now by definition
$T$ is not an alternative anti-Wick quantization if and only if
$$
T_{g^* h} \ne T_{g^*} T_{h}  
$$
for some $g, h \in \mathcal{P}$.
These two statements give the result.
$\quad \blacksquare$

\begin{corollary}
Suppose that 
there exists an element in $\mathcal{P}$ which  does not commute 
with some element in $\mathcal{P}^*$ and that $T$ is a monomorphism.
Then $T$ is not an alternative anti-Wick quantization.
\end{corollary}
\textbf{Proof:}
By hypothesis there exist elements $g, h \in \mathcal{P}$ such that $g^* h \ne h g^*$.
Since $T$ is a monomorphism, this implies that
$T_{g^* h } \ne T_{h g^*} $.
And now the previous theorem applies.
$\quad \blacksquare$

\vskip 0.4cm \noindent 
I suppose that the Toeplitz quantization of some non-commutative
algebras can be an alternative anti-Wick quantization,
but I have not constructed an example.
The results just presented indicate where not to look for such an example.
This remains  an open, though relatively minor, problem.

\vskip 0.4cm 
Just for completeness we read into the record two more related definitions.
\begin{definition}
We say that $T$ is a {\em Wick quantization} if 
$$
T_{h g^*}= T_h T_{g^*} 
$$
for all $g, h \in \mathcal{P}$.
We say that $T$ is an  {\em alternative Wick quantization} if 
$$
T_{g^* h}  = T_h T_{g^*}  
$$
for all $g, h \in \mathcal{P}$.
\end{definition}
These definitions are not expected in any way at all to describe a typical
Toeplitz quantization.
Their value lies in the possibility that some other types of quantizations of non-commutative 
algebras may have these properties.
If the range of the Toeplitz quantization $T$ consists of operators
which commute among themselves, then $T$ is trivially a Wick quantization.
Of course, this condition on $T$ is not what one wants in a quantum theory
and should be considered as a pathological condition.

\section{Canonical Commutation Relations}
\label{CCR-section}

We next want to consider the canonical commutation relations satisfied by these
creation and annihilation operators. 
The method of this section can be applied to all the examples in \cite{part2} and
\cite{quantum-plane} without further ado. 
We shall do this later on for one of the examples from \cite{quantum-plane}. 
One upshot of such an exercise is that the deformation parameter $q$ in those 
papers is seen to be independent of Planck's constant $\hbar$. 

Now our approach here is quite the opposite of the usual approach in which 
one starts with some generalization or modification of the standard canonical commutation relations
(considered as formal relations to be satisfied), 
and then one looks for realizations (namely, representations) of them as actual operators
in some Hilbert space. 
Here we would like to find the appropriate canonical commutation relations that arise from
a given Toeplitz quantization, that is, the operators are given first.

Our first observation is that the creation and annihilation operators all sit inside the
algebra $\mathcal{L}$.
So they generate a sub-algebra of $\mathcal{L}$, which is an object well known
in mathematical physics.
\begin{definition}
The sub-algebra of $\mathcal{L}$ generated by all the creation operators
$T_g$, where $g \in \mathcal{P}$, and all the annihilation operators $T_h$,
where $h \in \mathcal{P}^*$, is called the {\rm algebra of canonical commutation relations (CCR)}
and is denoted by $\mathcal{CCR}(\mathcal{A}, \mathcal{P})$.
\end{definition}
Alternatively, we will write 
$\mathcal{CCR}$ if context
resolves the ambiguity in this notation.
This may be a good time to point out that the inner product has been suppressed from our
notation of Toeplitz operators.
Therefore even the notation $\mathcal{CCR}(\mathcal{A}, \mathcal{P})$ is ambiguous.

We also define the \textit{Toeplitz algebra}, denoted $\mathcal{T}$, to be the sub-algebra
of $\mathcal{L}$ generated of all the Toeplitz operators $T_g$
for arbitrary symbols $g \in \mathcal{A}$.
Clearly, $\mathcal{CCR} \subset \mathcal{T}$.
An explicit description of either of the algebras $\mathcal{CCR}$ or
$\mathcal{T}$ seems to be in no way trivial in general.

Notice that in this abstract approach we first define the algebra of CCR before defining
the canonical commutation relations themselves.
Typically in studies in physics and mathematical physics, one defines the algebra of CCR
in terms of a presentation of generators and relations, where the relations are exactly the
canonical commutation relations.
In the present abstract approach this corresponds to writing $\mathcal{CCR}$ as the quotient of
some other free algebra $\mathcal{F}$ and then identifying the kernel 
of the quotient map $\pi: \mathcal{F} \to \mathcal{CCR}$  as the ideal of relations. 
Then we could pick a minimal set of generators of this ideal of relations as the CCR
of this theory. 
However, the trick is to do this (or at least some of it) in a functorial way, because otherwise
we will not have a general theory.

We propose the following construction. 
We define $\mathcal{F}$ to be the free algebra over $\mathbb{C}$ generated by
the set $\mathcal{P} \cup \mathcal{P}^*$.
Since $\mathcal{P} \cup \mathcal{P}^* \subset \mathcal{A}$,
we distinguish the product in $\mathcal{F}$ from that in $\mathcal{A}$ by writing
the algebra generators of $\mathcal{F}$ as $G_f$ for $f \in \mathcal{P} \cup \mathcal{P}^*$.
So $\mathcal{F}$ is the complex vector space with a basis given by all monomials 
$G_{f_1} G_{f_2} \cdots G_{f_n}$ with each $f_j \in \mathcal{P} \cup \mathcal{P}^*$.
The  algebra morphism 
$\pi: \mathcal{F} \to \mathcal{CCR}$ is defined on the algebra
generators of $\mathcal{F}$ by 
$\pi : G_f \mapsto T_f$ for all $f \in \mathcal{P} \cup \mathcal{P}^*$.
Since the algebra $\mathcal{F}$ is free on these generators, this defines $\pi$ uniquely.
Also since the elements $T_f$ for $f \in \mathcal{P} \cup \mathcal{P}^*$ are algebra
generators for the algebra $\mathcal{CCR}$, it follows that $\pi$ is surjective.
Moreover, $\pi (G_{f_1} G_{f_2} \cdots G_{f_n}) = T_{f_1} T_{f_2} \cdots T_{f_n}$
gives the map $\pi$ on a basis of $\mathcal{F}$.

\begin{definition}
Let $\pi: \mathcal{F} \to \mathcal{CCR}$ be as above.
Then we define the ideal of {\rm canonical commutation relations} (CCR) in $\mathcal{F}$ 
to be $\mathcal{R}:= \ker \pi$.
\end{definition}

This seems to be as far as one can go before getting down to the details of
picking ideal generators of $\mathcal{R}$.
It appears to be impossible to do that next step in a functorial way in this general setting. 
But it is reasonable to say that any minimal set of algebra generators of $\mathcal{R}$ is
a set of CCR.
Notice that such a set need not be unique in general.
So we are still some ways from having the typical situation found in most studies of CCRs.

We now discuss the well known, standard CCRs of quantum mechanics
in $\mathbb{R}^n$ in this setting. 
These are given by the generators $A_1, \dots A_n, A_1^\dag, \dots , A_n^\dag$
together with the relations (the standard CCR):
\begin{align}
     &A_j A_k - A_k A_j  \label{A-ccr}
     \\
     &A_j^\dag A_k^\dag - A_k^\dag A_j^\dag   \label{Adag-ccr}
     \\
     &A_j A_k^\dag - A_k^\dag A_j -  \delta_{j,k} \, \hbar \,1 \quad \mathrm{(Kronecker~delta)} 
     \label{A-Adag-qccr}
\end{align}
for $j, k \in \{ 1 , 2 , \dots , n\}$.
Here $\hbar >0$ is Planck's constant.
Notice that we have deliberately written these as relations to be quotiented out
and that we have not used the standard notation for commutators.
One important point, often sluffed over, is that (\ref{A-ccr}) and (\ref{Adag-ccr})
are non-trivial relations mathematically, since they impose the commutativity of certain pairs of
generators in the quotient algebra of CCR. 
 The corresponding generators of the free algebra do not commute, of course.
 However, in physics the intuition is that commuting operators are like objects in classical
 mechanics and so are deemed to be trivial in the quantum setting. 
 Using this physics intuition, the only non-trivial case is (\ref{A-Adag-qccr})
 when $j=k$, while from a mathematical perspective all the cases 
 of (\ref{A-Adag-qccr}) as well as (\ref{A-ccr}) and (\ref{Adag-ccr}) are non-trivial.

Instead of concerning ourselves with 
what is trivial and what is not (and from whose point of view), let us simply
note that the relations (\ref{A-ccr}), (\ref{Adag-ccr}) and, when $j \ne k$ (\ref{A-Adag-qccr}),
are homogeneous elements (in this case of degree~$2$) in the free algebra, while
(\ref{A-Adag-qccr}) for $j = k$ is not a homogeneous element.
Also, Planck's constant only plays a role in (\ref{A-Adag-qccr}) for $j = k$
and then only in a lower order term.

The free algebra $\mathcal{F}$ introduced above is also a graded algebra, where the linear span
of all of the
basis elements $G_{f_1} G_{f_2} \cdots G_{f_n}$ for some fixed integer $n \ge 0$ and
$f_j \notin \mathbb{C} 1$ for $j = 1, \dots , n$ 
is by definition the subspace of homogeneous elements of degree $n$. 
The identity element $1 \in \mathcal{F}$ has degree zero. 

\begin{definition}
Let $\mathcal{R} \subset \mathcal{F}$ be the ideal of CCR as above. 
Any homogeneous element in $\mathcal{R}$ is called a {\rm classical relation}
while any non-homogeneous element in $\mathcal{R}$ is called a {\rm quantum relation}. 
\end{definition}

This dichotomy is important more for ideal generators of $\mathcal{R}$ rather than for arbitrary
 elements in $\mathcal{R}$ itself.
 For example, using this dichotomy, one sees that the $q$-commutation relation
 $A A^\dag - q A^\dag A$ (usually written as $x y - q y x$) 
 for $q \in \mathbb{C}$ is a classical relation, 
 while the relation $A A^\dag - q A^\dag A- \hbar 1$
is a quantum relation.
Notice that both of these relations,
classical and quantum, arise in the study of Toeplitz operators
associated with the quantum plane.
See \cite{quantum-plane}.

The next definition is motivated by the examples discussed above.
\begin{definition}
Let $R \in \mathcal{R}$ be a non-zero relation. 
Then we can write $R$ uniquely as $R = R_0 + R_1 + \cdots + R_n$,
where $\deg \, R_j = j$ for each $j = 0,1, \dots, n$ and $R_n \ne 0$.
Then we say that $R_n$ is the {\rm classical relation associated to~$R$}.
\end{definition}
Notice that $R_n$ is indeed a classical relation.
Both of the cases $R_n \in \mathcal{R}$ and $R_n \notin \mathcal{R}$ can occur.
Intuitively, to get the classical relation $R_n$ from $R$ we throw away the `quantum corrections' 
$R_0, R_1, \dots, R_{n-1}$ in $R$.
We let 
$$
\mathcal{R}_{cl}:= \langle R_n \, | \, R_n 
\mathrm{~is~the~classical~relation~associated~to~some~}
R \rangle,
$$
where $R$ ranges over some set of generators of $\mathcal{R}$ and
the brackets $\langle \cdot \rangle$ indicate that we are taking the
two-sided ideal in $\mathcal{F}$ generated by the elements inside the brackets.
\begin{definition}
\label{define-DQ-algebra}
The {\rm dequantized algebra} associated to $\mathcal{A}$ is defined to be
$$
\mathcal{DQ}:= \mathcal{F} / \mathcal{R}_{cl}.
$$
\end{definition}
Note that $\mathcal{DQ}$ need not be commutative. 

We can realize $\mathcal{DQ}$ as the case $\hbar =0$ of a family of algebras
defined for all $\hbar \in \mathbb{C}$ and with $\hbar =1$ corresponding to $\mathcal{CCR}$.
Of course, when $\hbar > 0$ we interpret $\hbar$ as Planck's constant.
To achieve this we define the $\hbar$-deformed relations to be
\begin{align}
\mathcal{R}_\hbar :=&\langle   \hbar^{n/2} R_0 + \hbar^{(n-1)/2} R_1 + \cdots + 
\hbar^{1/2} R_{n-1} + R_n \rangle
 \label{define-Rhbar-1x}
\\
= &\langle R_0 + \hbar^{-1/2} R_1 + \cdots + 
\hbar^{-(n-1)/2} R_{n-1} + \hbar^{-n/2} R_n   \rangle,
   \label{define-Rhbar-2x}
\end{align}
using the above notation $R = R_0 + R_1 + \cdots + R_n$, where 
$R$ ranges over a set of generators of $\mathcal{R}$.   
And then we define 
$$
    \mathcal{CCR}_\hbar :=  \mathcal{F} / \mathcal{R}_{\hbar}.
$$
In the second expression (\ref{define-Rhbar-2x}) the powers of $\hbar^{-1/2}$ correspond
to the degree of homogeneity of each of the terms, while in the first expression (\ref{define-Rhbar-1x}) 
each of the homogeneous terms has been given its intuitively correct 
degree of `quantumness'.
The first expression (\ref{define-Rhbar-1x}) also clarifies in a formal way what happens when 
we take the limit when $\hbar \to 0$. 
For $\hbar \ne 0$ the two expressions  (\ref{define-Rhbar-1x}) and  (\ref{define-Rhbar-2x})
are clearly equivalent, but for $\hbar =0$ only the definition (\ref{define-Rhbar-1x}) makes sense.
In physics one considers $\hbar > 0$, but here the discussion is valid for $\hbar \in \mathbb{C}$.

An example of this quantization scheme is given in 
\cite{quantum-plane} in the setting of the quantum plane of Manin. 
In the notation of that paper the annihilation operator is $A = T_{ \overline{\theta} }$ 
and the creation operator is $A^\dagger = T_\theta$. 
In a special case described there in detail (see the discussion related to 
Eq.~(27) in \cite{quantum-plane}), 
one has the commutation relation
$$
          [ A, A^\dagger ]_{q^{-1}} = A A^\dagger - q^{-1} A^\dagger A = I,
$$
where $q \in \mathbb{C} \setminus \{ 0 \}$. 
The occurrence of $q^{-1}$ in this commutation relation arises because
we are following the notational conventions of \cite{quantum-plane}. 
Introducing Planck's constant $\hbar$ according to the above procedure gives 
us the relation 
$$
[ A, A^\dagger ]_{q^{-1}} = \hbar I.
$$
This clearly demonstrates, as mentioned earlier, that $q$ and $\hbar$ 
are independent parameters. 
This in turn gives the standard canonical commutation relation when 
one lets $q=1$. 
The corresponding classical relation is $[ A, A^\dagger ]_{q^{-1}} = 0$. 
Therefore, we have that $A A^\dagger = q^{-1} A^\dagger A$ holds
in $\mathcal{DQ}$. 
This shows that the motivating example in (\ref{A-Adag-qccr}) for the 
physically non-trivial case $j=k$ is recovered here when $q=1$ and $n=1$. 
Generalizing the results in \cite{quantum-plane} to the case $n \ge 2$ 
is straightforward, thereby allowing us to recover (\ref{A-Adag-qccr})  as well
for $j=k$, $q=1$ and $n \ge 2$. 

Again we remark that the dequantized algebra  $\mathcal{DQ}$ 
in Definition~\ref{define-DQ-algebra}
could be non-commutative 
even though intuitively one feels that
the construction of the dequantized algebra is some sort of \textit{dequantization}.
Another blow to intuition can occur in the quantization process too, 
since the ideal $\mathcal{R}$ 
could logically speaking be generated by a set of classical generators. 
In that case Planck's constant $\hbar$ plays no role in the definition of $\mathcal{R}_\hbar $. 
Actually, in such a case one would have  $\mathcal{R}_\hbar = \mathcal{R} = \mathcal{R}_{cl}$. 
In other words, the result of the quantization in such a case would not be a quantum theory 
from a physical point of view. 
However, I have so far found no examples of these logically possible,
counter-intuitive situations. 
Another quite likely possibility is that $\mathcal{R}= 0$. 
In that case we would also have $\mathcal{R}_\hbar = \mathcal{R}_{cl}=0$.

We have included Planck's constant $\hbar$ to emphasize that this theory has 
semi-classical behavior (more precisely, what happens when $\hbar$ tends to zero) as well
as a classical counterpart $\mathcal{DQ}$ (that is, what happens when we put $\hbar$ equal to zero).
However, the semi-classical theory as well as
the general relation between the algebras $\mathcal{A}$ and $\mathcal{DQ}$ remain
as open problems.
Each of the algebras $\mathcal{A}$ and $\mathcal{DQ}$ is a `classical' algebra 
(though possibly in different senses of the word `classical')
with $\mathcal{CCR}$ being an intermediate quantum algebra
of interest.
The Toeplitz algebra $\mathcal{T}$ is also a quantum algebra with its own intrinsic interest.

\section{Toeplitz Operators with Symbols in $SU_q(2)$} 
\label{SUq2-section}

In this section we give an example of 
Toeplitz operators whose symbols are in $\mathcal{A} = SU_q(2)$, a 
real form of $SL_q(2)$. 
These operators act on the Manin quantum plane,
which is realized as a dense sub-space of a Hilbert space. 
(See \cite{manin}, p.~131 where the notation $A^{2 | 0}_q$ is used for
this quantum plane.) 
As far as I am aware, this is the first example of Toeplitz operators 
defined with symbols in $SU_q(2)$, though for the well studied case $q=1$ 
it is difficult to imagine that this has not been done before, though perhaps
is a disguised form. 
Anyway, 
this example leads me to conjecture that similar examples also exist for 
the $q$-deformations of other compact Lie groups. 

This example shows that none of the definitions of the $*$-operation 
on $SL_q(2)$, the inner
product on $\mathcal{A}$ and the sub-algebra $\mathcal{P}$ is unique. 
The moral, as indicated earlier, is that we need more structure 
than the algebra itself in order to quantize the algebra $SL_q(2)$. 

First, we review in this paragraph some known properties of $SU_q(2)$. 
There are many fine references for this material, one of which is \cite{timmerman}. 
We define $SU_q(2)$ as the universal $*$-algebra over $\mathbb{C}$ 
on the generators $a$ and $c$
satisfying these relations:
\begin{align*}
      &a c = q c a \qquad a c^* = q c^* a 
      \qquad 
      c c^* = c^* c
      \\
      &a^* a + c^* c = 1 \qquad  a a^* + q^2 c^* c = 1,
\end{align*}
where $q \in \mathbb{R} \setminus \{ 0 \}$. 
This space has the structure of a $*$-Hopf algebra, but for now
we consider it just as a $*$-algebra. 
This vector space has a useful basis, which we now discuss.
For $k \in \mathbb{Z}$ and $l,m \in \mathbb{N}$ 
we define
$$
      \varepsilon_{k l m} := 
      \left\{ 
      \begin{array}{ll}
      a^k \, c^l \, (c^*)^m = q^{k m} \, (c^*)^m \, a^k \, c^l & \mathrm{~for~} k \ge 0,
      \\
      \\
      (a^*)^{-k} \, c^l \, (c^*)^m = (a^*)^{-k} \, (c^*)^m \,  c^l  & \mathrm{~for~} k < 0.
      \end{array}
      \right.
$$
Then it is known that the set 
$\{  \varepsilon_{klm} ~|~ k \in \mathbb{Z}, \, l,m \in \mathbb{N}  \}$
is a vector space basis of $SU_q(2)$. 

To define Toeplitz operators we need to define a sub-algebra 
(but not a sub-$*$-algebra) of $\mathcal{A} = SU_q(2)$. 
So, to achieve this we define $\mathcal{P}$ to be the sub-algebra generated by the
elements $a$ and $c$. 
Since $a c = q c a$, this is a copy of a Manin quantum plane. 
It has a Hamel basis given by $ \{ a^i c^j ~|~ i, j \in \mathbb{N} \}$. 
Also, the expressions given above for $ \varepsilon_{k l m}$ show that 
$\mathcal{A} = \mathcal{P}^* \mathcal{P}$. 
Finally, notice that $\mathcal{P}^*$ is a sub-algebra isomorphic to a Manin quantum plane 
and that $\mathcal{P} \cap \mathcal{P}^* = \mathbb{C}1$. 
So, $a$ and $c$ can be viewed as holomorphic variables while $a^*$ and $c^*$ 
as anti-holomorphic variables. 

We also need an inner product on $\mathcal{A} = SU_q(2)$. 
We first define this on pairs of basis vectors and then extend to the unique
sesquilinear form on $SU_q(2)$. 
We use the convention that {\em sesquilinear} means anti-linear in the first 
entry and linear in the second. 
So for $k, r \in \mathbb{Z}$ and $l, m, s, t \in \mathbb{N}$ we define
$$
    \langle \varepsilon_{k l m} , \varepsilon_{r s t} \rangle_{\mathcal{A}} 
    := w (l + t) \, \delta_{k, r} \, \delta_{l+t, m+s}. 
$$ 
Here $w : \mathbb{N} \to (0, \infty)$ is any strictly positive, real function whose values are 
called {\em weights}.
Also, we are using the standard notation for the Kronecker delta. 
This definition is motivated by similar definitions in the references \cite{part1}, 
\cite{quantum-plane} and originally in \cite{bargmann}. 
This choice for the inner product is not unique. 
The theory 
goes through just fine  with appropriate changes for other choices. 

Whether this inner product is non-degenerate is a question that need not be 
considered for now. 
What is important in this paper is its restriction to the sub-algebra $\mathcal{P}$. 
So for $k, l, r, s \in \mathbb{N}$ we have 
$$
     \langle a^k c^l , a^r c^s \rangle_{\mathcal{A}}  = 
     \langle \varepsilon_{k, l, 0} , \varepsilon_{r, s, 0} \rangle_{\mathcal{A}}  =
      w (l) \, \delta_{k, r} \, \delta_{l, s} =
      w (s) \, \delta_{k, r} \, \delta_{l, s}
$$
This shows that the inner product restricted to $\mathcal{P}$ is
positive definite. 
So, we define 
$$
\varphi_{k l}:=  (w(l))^{-1/2} \varepsilon_{k, l, 0} =  (w(l))^{-1/2} a^k c^l. 
$$ 
Then $\Phi := \{ \varphi_{k, l} ~|~ k,l \in \mathbb{N} \}$ 
is an orthonormal Hamel basis of $\mathcal{P}$. 
The completion of the pre-Hilbert space $\mathcal{P}$ with respect to
this positive definite inner product is denoted by $\mathcal{H}$, a
Hilbert space. 
It follows that $\Phi$ is an orthonormal basis of $\mathcal{H}$. 
We have one remaining condition to check, namely Condition~3 which we repeat 
in this context:

\vskip 0.4cm \noindent
3. For every $f \in \mathcal{A}$, the set 
$\Phi_f = \{ \varphi_{i,j} \in \Phi \, | \, \langle \varphi_{i,j} , f \rangle_{\mathcal{A}} \ne 0 \}$
is finite.

\vskip 0.4cm \noindent
We first consider the case when $f = \varepsilon_{klm}$ for given values 
$k \in \mathbb{Z}$ and $l,m \in \mathbb{N}$. 
Then for integers $i, j \ge 0$ we have 
$$
        \langle \varphi_{i,j} , \varepsilon_{klm} \rangle_{\mathcal{A}}  = 
         \langle (w(j))^{-1/2} \varepsilon_{i, j, 0}  , \varepsilon_{klm} \rangle_{\mathcal{A}}  = 
         (w(j))^{-1/2} w(j+m) \, \delta_{j+m, l} \, \delta_{i, k}. 
$$
This inner product is non-zero if and only if $k = i \ge 0$ and $l-m = j \ge 0$. 
Thus we see that there is at most
one solution $i,j \in \mathbb{N}$ for $k, l, m$ as given above. 
Therefore, the set $\Phi_f$ is finite provided that $f = \varepsilon_{klm}$. 
But any $f  \in \mathcal{A}$ is equal to a finite linear combination of the $\varepsilon_{klm}$'s, 
since these form a basis. 
So, a necessary condition for 
$\langle \varphi_{i,j} , f \rangle_{\mathcal{A}} \ne 0 $ is that 
\begin{equation}
\label{necessary-condition}
\langle \varphi_{i,j} , \varepsilon_{klm} \rangle_{\mathcal{A}}  \ne 0
\end{equation}
for at least 
one of the $\varepsilon_{klm}$'s appearing in that finite linear combination. 
Hence, there are only finitely many (possibly zero) 
pairs $i,j$ given a specific $f \in \mathcal{A}$ such that (\ref{necessary-condition}) hold.
We conclude that Condition~3 is satisfied. 

We now have all the ingredients needed for defining the projection 
operator $P_K : \mathcal{A} \to \mathcal{P}$. 
We recall that the formula in this setting becomes 
\begin{equation}
   P_K f = \sum_{i,j \ge 0} \langle   \varphi_{i,j} , f \rangle_{ \mathcal{A} } \,\, \varphi_{i,j} 
\end{equation}
for all $ f \in \mathcal{A}$.  
Moreover, the Toeplitz operator associated to the symbol $g \in \mathcal{A} = SU_q(2)$
is $ T_g = P_K M_g$ as in the general theory. 
Also, $T_g : \mathcal{P} \to \mathcal{P}$, that is it maps the Manin quantum plane to itself,
and is a densely defined linear operator in the Hilbert space $\mathcal{H}$. 
Recall that $M_g$ is multiplication from the right by $g$. 

The creation operators associated to the two $*$-algebra generators $a$ and $c$
are $T_a = M_a$ and  $T_c = M_c$. 
These both raise the degree by $1$ when acting on homogeneous elements 
of $\mathcal{P}$. 
And each of these homogeneous elements is a multiple of a
basis element $\varphi_{i, j}$. 
Explicit calculations easily give 
$$
        T_a (\varphi_{i,j}) = q^{-j} \varphi_{i+1,j} 
        \qquad \mathrm{and} \qquad
        T_c (\varphi_{i,j}) = \left( \dfrac{w(j+1)}{w(j)} \right)^{1/2} \varphi_{i,j+1}. 
$$
Specifically, $T_a$ has bi-degree $(1,0)$ in the variables $a,c$ and
$T_c$ has bi-degree $(0,1)$ in $a,c$. 
A curious fact here is that the formula for $T_a$ does not involve the weights, 
while that for $T_c$ does not involve $q$. 
These identities in turn immediately imply the $q$-commutation relation
$$
       [ T_c,  T_a ]_q = T_c T_a - q \, T_a T_c = 0. 
$$

The corresponding annihilation operators $T_{a^*}$ and $T_{c^*}$ are degree $-1$ 
linear maps on homogeneous elements. 
Explicit formulas, proved below, are given for $i,j \ge 0$ by
\begin{align*}
   T_{a^*} (\varphi_{i,j}) &= q^j \, 
   \left( 
       1 - q^2 \dfrac{w(j+1)}{w(j)}
   \right) 
   \, \varphi_{i-1,j}
   \\
   T_{c^*} (\varphi_{i,j}) &=  
   \left(
        \dfrac{w(j)}{w(j-1)}
    \right)^{1/2} 
   \varphi_{i,j-1}, 
\end{align*}
where the right side of either identity is taken to be $0$ if one of the sub-indices
of $\varphi$ is $-1$. 
Again, the bi-degrees with respect to $a,c$ are as expected: 
$(-1,0)$ for $T_{a^*}$ and $(0,-1)$ for $T_{c^*}$. 

However, with our choice of inner product 
the annihilation operators are not necessarily the adjoints of
the creation operators as we shall see a little later on. 
It is an open problem to find another inner product for which 
the annihilation operators are the adjoints of
the creation operators and everything else works out well. 

So, we compute the annihilation operators directly from the definition. 
In the following we use $\langle \cdot , \cdot \rangle$ to mean
$\langle \cdot , \cdot \rangle_{\mathcal{A}}$.  
For example, to get the formula for $T_{a^*} $ we start as follows:
\begin{align}
    T_{a^*} (\varphi_{i j}) &= P_K (\varphi_{i j} a^* ) \nonumber
    \\
    &= w(j)^{-1/2} \sum_{k, l \ge 0} \langle \varphi_{k l} , a^i c^j a^* \rangle \varphi_{k l} \nonumber
    \\
    \label{T-sub-c-star}
    &= w(j)^{-1/2} \sum_{k, l \ge 0} w(l)^{-1/2} 
    \langle \varepsilon_{k, l, 0} , q^j a^i a^* c^j \rangle \varphi_{k l}.
\end{align}
As before, this is valid for $i,j \ge 0$. 
At this point, we see that the case $i=0$ leads to
\begin{align*}
    T_{a^*} (\varphi_{0, j}) &= w(j)^{-1/2} \sum_{k, l \ge 0} w(l)^{-1/2} q^j
    \langle \varepsilon_{k, l, 0} ,  a^* c^j \rangle \varphi_{k l}
    \\
    &= w(j)^{-1/2} \sum_{k, l \ge 0} w(l)^{-1/2} q^j
    \langle \varepsilon_{k, l, 0} ,  \varepsilon_{-1, j, 0} \rangle \varphi_{k l} 
    \\
    &= w(j)^{-1/2} \sum_{k, l \ge 0} w(l)^{-1/2} q^j w(l) \, \delta_{k, -1} \, \delta_{l, j} 
    \, \varphi_{k l} 
    \\
    &= 0.
\end{align*}
So, we consider the remaining case $i \ge 1$. 
Then we have to consider this expression which appears in (\ref{T-sub-c-star}): 
\begin{align*}
          \langle \varepsilon_{k, l, 0} , q^j a^i a^* c^j \rangle 
          &= q^j \langle \varepsilon_{k, l, 0} ,  a^{i-1} a a^* c^j \rangle
          \\
          &=  q^j \langle \varepsilon_{k, l, 0} ,  a^{i-1} (1 - q^2 c^* c) c^j \rangle
          \\
          &= q^j 
          \left(
               \langle \varepsilon_{k, l, 0} , a^{i-1} c^j \rangle
               -
                q^2 \langle \varepsilon_{k, l, 0} ,  a^{i-1} c^{j+1} c^* \rangle
          \right)
          \\
          &= q^j 
          \left(
               \langle \varepsilon_{k, l, 0} , \varepsilon_{i-1, j, 0} \rangle
               -
                q^2 \langle \varepsilon_{k, l, 0} ,  \varepsilon_{i-1, j+1, 1}  \rangle
          \right)
          \\
          &= q^j  \left( w(l)  \delta_{k, i-1} \delta_{l,j} - q^2 w(l+1) \delta_{k, i-1} \delta_{l+1,j+1} \right)
          \\
          &= q^j   \left( w(l) - q^2 w(l+1) \right) \delta_{k, i-1} \delta_{l,j}. 
\end{align*}
Substituting this back into (\ref{T-sub-c-star}) we now obtain 
\begin{align*}
        T_{a^*} (\varphi_{i, j}) &=  
         w(j)^{-1/2} \sum_{k, l \ge 0} w(l)^{-1/2} 
    \langle \varepsilon_{k, l, 0} , q^j a^i a^* c^j \rangle \varphi_{k l}
    \\
    &= w(j)^{-1/2} \sum_{k, l \ge 0} w(l)^{-1/2} 
    q^j  \left( w(l) - q^2 w(l+1) \right)
    \delta_{k, i-1} \delta_{l,j} 
    \, \varphi_{k l}
    \\
    &= w(j)^{-1/2} 
    w(j)^{-1/2} 
    q^j  \left( w(j) - q^2 w(j+1) \right) 
    \, \varphi_{i-1, j}
    \\
    &= q^j \, \dfrac{ w(j) - q^2 w(j+1)}{w(j)}   \,\, \varphi_{i-1, j}. 
\end{align*}

For $T_{c^*}$ the calculation is much simpler. 
\begin{align*}
          T_{c^*} (\varphi_{i j}) &= P_K (\varphi_{i j} c^* ) 
    \\
    &= w(j)^{-1/2} \sum_{k, l \ge 0} \langle \varphi_{k l} , a^i c^j c^* \rangle \, \varphi_{k l} 
    \\
    &=   w(j)^{-1/2} \sum_{k, l \ge 0} w(l)^{-1/2} \langle  \varepsilon_{k,l,0}, 
    \varepsilon_{i ,j, 1} \rangle \, \varphi_{k l} 
    \\
    &=   w(j)^{-1/2} \sum_{k, l \ge 0}  w(l)^{-1/2} w(l+1) \delta_{k, i} \, \delta_{l+1, j} 
    \, \varphi_{k l} 
    \\
    &=   w(j)^{-1/2} w(j-1)^{-1/2} w(j) 
    \, \varphi_{i, j-1}
    \\
    &= 
    \left(
        \dfrac{w(j)}{w(j-1)}
    \right)^{1/2} 
    \varphi_{i, j-1}, 
\end{align*}
provided that $j \ge 1$ and $i \ge 0$. 
Clearly, this argument also shows that 
$ T_{c^*} (\varphi_{i, 0}) = 0$.
So we see that the formula for 
$T_{c^*}$ does not depend on $q$, but does depend on 
the weights. 

Next, we include for the record the formulas for the 
adjoints of the two creation operators 
(as linear operators on the pre-Hilbert space $\mathcal{P}$)
considered above:
$$
  T_a^* (\varphi_{i j} ) =  q^{-j} \varphi_{i-1 , j} \quad \mathrm{and} \quad   
  T_c^* (\varphi_{i j} ) = \left( \dfrac{w(j)}{w(j-1)} \right)^{1/2} \varphi_{i,j-1}. 
$$
These are easy enough to check out and so are left to the reader.
Note that, as previously mentioned, this shows that $T_{a^*}$ 
is not equal in general to $T_a^*$. 
However, it does turn out that $T_{c^*}$ is equal to $T_c^*$ on 
the domain $\mathcal{P}$.
This asymmetry in the roles of $a$ and $c$ follows, of course, 
from the asymmetries in the defining relations for these elements and their
conjugates. 
Nonetheless, there should be a deeper understanding of the 
structure of this asymmetry. 

This example could be augmented with more formulas for the
creation operators $T_{a^k c^l}$, the corresponding annihilation 
operators, the adjoints of all of these as well as all possible commutation 
relations among these operators. 
For now, we content ourselves with just some commutation relations. 
For example, the $q$-commutation relation for the adjoint operators is
$$
      [ T_a^* , T_c^* ]_q = 0. 
$$
Mixing creation operators with their adjoints operators we easily calculate that
\begin{align}
      [ T_a^*, T_a ] &= 0, \nonumber
      \\
     [ T_c , T_a^* ]_q &= 0,  \nonumber
      \\
     [ T_a , T_c^* ]_q &= 0,  \nonumber
     \\
     \label{four-mixed}
      [ T_c^*, T_c]_q \, \varphi_{i,j} &= \kappa_j \, \varphi_{i,j}, 
\end{align}
where
$$
      \kappa_j =  \dfrac{w(j+1)}{w(j)}  - q \dfrac{w(j)}{w(j-1)}. 
$$
So, in general, the $q$-commutator $ [ T_c^*, T_c]_q$ is
diagonalized in the basis $\{  \varphi_{i,j}  \}$ and the exact 
eigenvalues $\kappa_j$ depend on $q$ and the weights. 
We can clean up the formula by defining $K$ to be that particular 
diagonalized operator, in which case we get $ [ T_c^*, T_c]_q =K$. 
If $K = I$, which does occur for appropriate choices of the weights, 
then we get a `standard' $q$-commutation relation whose quantization 
is $ [ T_c^*, T_c]_q = \hbar I$. 
The first three commutation relations in (\ref{four-mixed}) 
are homogeneous relations of degree $2$, 
and so they are classical relations which therefore 
remain unchanged under quantization. 
In a colloquial manner of speaking one could say that $a$ is a `classical' variable
and that $c$ is a `quantum' variable. 
This again indicates an asymmetry between $a$ and $c$. 
We remind the reader that both $a$ and $c$ are holomorphic variables.

\section{Concluding Remarks}
\label{CRemarks-section}

There are many papers, especially in the physics literature,
dedicated to the study of a given deformation of the canonical commutation relations (CCRs).
Realizations of these deformed CCRs can be a non-trivial problem.
Often the solution is given not by using a Toeplitz quantization but rather some other approach. 
Also the approach in those papers typically involves the definition of the
algebra under study in terms of generators and relations. 
This leads to highly specific studies of rather concrete mathematical structures.
A `slight' change of the presentation in terms of generators and relations can
entail a rather different theory. 
Also, one is faced with the often intractable problem of identifying when 
two presentations in terms of generators and relations define isomorphic objects.

As mentioned earlier the approach in this paper is quite the opposite.
Here we start with the Toeplitz quantization of an algebra and then look for 
the corresponding generalized CCR's.
And we have not imposed many restrictions on the algebra $\mathcal{A}$ besides the quite standard
ones of associativity and existence of a unit. 
Also we require a $*$-operation and an inner product.
We have avoided the use of generators and relations as a starting point.
Of course, one can generalize or modify any theory, and for this theory 
one could drop the associativity condition or the existence of the unit. Or these could be
replaced by other conditions.
Similar comments apply to the $*$-operation and the inner product.
While these are possibilities for further research, we think that the theory as presented
here is still quite rudimentary and merits further study.
For example, we look forward to an understanding of how to find 
the (best?) generators of the generalized CCR's associated with a given Toeplitz quantization.

Finally, here are some comments on other Toeplitz quantizations
which use non-commuting symbols.
First there is the impressive monograph \cite{BandS} by B\"ottcher and Silbermann.
These authors, and the researchers associated with them, have produced a significant body of 
work on Toeplitz operators whose symbols are matrices 
with entries in various function spaces or in an algebra.
In this regard also see the papers \cite{karlovich1} and \cite{karlovich2} by Karlovich.
These works can hardly be described in a few words, but it seems that they always use measures
and that their Toeplitz operators act on functions, albeit vector-valued functions.
Their works include the study of Toeplitz operators in Banach spaces, such as $L^p$ and $H^p$.
Of course, in the present paper we  do not use measures but we do use an inner product. 
And our Toeplitz operators are only defined in a Hilbert space.
A major difference in emphasis is that the B\"ottcher-Silbermann school 
takes an operator theory approach,
whereas we are also treating topics because of their interest in physics 
and npn-commutative geometry as well as in analysis and operator theory.

The papers \cite{ali-englis1} and \cite{ali-englis2} by Ali and Englis use matrix valued symbols.
So again, these symbols are functions but with values in a non-commutative algebra. 
Their results are in the setting of $L^2$ spaces, so there is a measure being used. 
Their papers are concerned with Berezin-Toeplitz quantization, where one has quantum Hilbert spaces
$\mathcal{H}_\hbar$ indexed by Planck's constant $\hbar > 0$. 
These two papers are concerned with the asymptotics as $\hbar \to 0$.
That is a mathematical-physics approach, but treats themes complementary of those of this paper.
The paper \cite{kerr} of Kerr is similar to the work of Ali and Englis,
but now the symbols are matrices with
entries in a scalar valued Bergman space.
So this is based on a measure, and it also has more of a flavor of functional analysis
and  operator theory.

The papers \cite{borthwick-et-al} by Borthwick et al. and \cite{iuliu} by
Iuliu-Lazaroiu et al. study super-Toeplitz operators, that is, those that arise naturally in
super-manifold theory.
The symbols are super-functions, meaning they have commuting and anti-commuting parts.
This theory arose from Berezin's work in quantum physics and has become a
research area in and of itself in geometry. 
However, we find it to be rather complementary to the current approach.

None of these prior works was known to me until I was finishing up this paper.
Those works may have superficial similarities to this paper, but are not sources for it.
A major, important feature of this paper is that it provides a quantization scheme without
using a measure, or some sort of generalization of a measure as is done in \cite{csq}.
And this is a significant difference of this paper from those mentioned above.
Also, this theory applies to a rather wide class of non-commutative algebras.
Finally, 
we are presenting a theory intended to be applicable 
in operator theory, in mathematical physics as well as in 
non-commutative geometry.

\vskip 0.6cm \noindent
\textbf{Acknowledgments:} 
I am most whole heartedly grateful to S.T.~Ali, 
J.~Cruz Sampedro, M.~\dju, 
 J.-P. Gazeau and R.A.~Mart\'inez-Avenda{\~n}o 
for their clarifying remarks and incisive questions. 

This paper is dedicated to the memory of Jaime Cruz Sampedro whose
great friendship and mathematical intelligence I will dearly miss.

\end{document}